\documentclass[aps,prb,twocolumn,superscriptaddress,showpacs]{revtex4-1}
\usepackage{hyperref}
\usepackage{graphicx}
\usepackage{revsymb4-1}
\usepackage{units}
\usepackage{miller}

\newcommand{\BFA}{BaFe$_{2}$As$_{2}$}
\newcommand{\SFA}{SrFe$_{2}$As$_{2}$}

\newcommand{\TN}{T$_N$}		% N\'eel temp
\newcommand{\Ts}{T$_s$}		% structural temp
\newcommand{\Tc}{T$_c$}		% critical temp 
\newcommand{\qvec}{$\mathbf{q}$}
\newcommand{\Qvec}{$\mathbf{Q}$}

\newcommand{\degree}[1]{$#1^\circ$}
\newcommand{\appr}{$\approx$}

\begin{document}
%\linenumbers

\title{Close correlation between magnetic properties and the soft phonon mode of the structural transition in \BFA\ and \SFA}
%\emph{A}Fe$_{2}$As$_{2}$ (\emph{A}=Ba,Sr)}

\author{D.~Parshall}
%	\affiliation{University of Colorado, Department of Physics, Boulder, CO 80309}
%	\affiliation{National Institute of Standards and Technology, NIST Center for Neutron Research, Gaithersburg, MD 20899}
	\affiliation{NIST Center for Neutron Research, Gaithersburg, MD 20899}
	\email{parshall@nist.gov}   %optional

\author{L.~Pintschovius}
	\affiliation{Karlsruhe Institut f\"ur Technologie, Institut f\"ur Festk\"orperphysik, P.O.B. 3640, D-76021 Karlsruhe, Germany}

\author{J.~L.~Niedziela}
	\affiliation{Oak Ridge National Laboratory, Instrument and Source Division, Oak Ridge, TN 37831}

\author{J.-P.~Castellan}
	\affiliation{Karlsruhe Institut f\"ur Technologie, Institut f\"ur Festk\"orperphysik, P.O.B. 3640, D-76021 Karlsruhe, Germany}
	\affiliation{Laboratoire L\'eon Brillouin, CEA-Saclay, F-91191 Gif-sur-Yvette Cedex, France}

\author{D.~Lamago}
	\affiliation{Karlsruhe Institut f\"ur Technologie, Institut f\"ur Festk\"orperphysik, P.O.B. 3640, D-76021 Karlsruhe, Germany}

\author{R.~Mittal}
	\affiliation{Solid State Physics Division, Bhabha Atomic Research Centre, Trombay, Mumbai, 400 085, India}

\author{Th.~Wolf}
	\affiliation{Karlsruhe Institut f\"ur Technologie, Institut f\"ur Festk\"orperphysik, P.O.B. 3640, D-76021 Karlsruhe, Germany}

\author{D.~Reznik}
	\affiliation{University of Colorado, Department of Physics, Boulder, CO 80309}

\date{\today}

\begin{abstract} 
Parent compounds of Fe-based superconductors undergo a structural phase transition from a tetragonal to an orthorhombic structure. We investigated the temperature dependence of the frequencies of transverse acoustic (TA) phonons that extrapolate to the shear vibrational mode at the zone center, which corresponds to the orthorhombic deformation of the crystal structure at low temperatures in \BFA\ and \SFA. We found that acoustic phonons at small wavevectors soften gradually towards the transition from high temperatures, tracking the increase of the size of slowly fluctuating magnetic domains. On cooling below the transition to base temperature the phonons harden, following the square of the magnetic moment (which we find is proportional to the anisotropy gap). Our results provide evidence for close correlation between magnetic and phonon properties in Fe-based superconductors.
\end{abstract}

%\pacs{74.70.Xa 75.30.Gw 64.70.K- 75.40.Gb}
%\keywords{}

\maketitle

%%%%%%%%%%%%%%%%%%%%%%%%%%%%%%%%%%%%%%%%%%%%%%%%%%%%%%%%%%%%%%%%%%%%%%%%%%%%%%%%
\section{Introduction}

The parent compounds of the ferropnictide superconductors are metallic but not superconducting, and order antiferromagnetically. Just like the cuprates, they become superconducting upon doping, which gradually suppresses  antiferromagnetic order.\cite{Dai2012}  There is strong coupling between the magnetic degrees of freedom and the lattice. The most important experimental evidence for this coupling is a structural transition from tetragonal to orthorhombic, which is thought to be driven by magnetism \cite{Yildirim09-1}. Further, it was reported that the arsenic A$_{1g}$ mode can generate antiferromagnetic (AFM) order dynamically \cite{KimKW2012}, and Raman scattering shows strong splitting of the in-plane E$_g$ mode in the ordered state \cite{Chauviere2009}. These observations have led several groups to suggest enhanced electron-phonon coupling via the spin channel \cite{Egami10-1,Yndurain2011}. 

On the other hand, density-functional theory (DFT) predicts only very weak electron-phonon coupling \cite{Boeri09}, which is not directly contradicted by any experimental result so far.  Previous work using inelastic x-ray scattering (IXS) \cite{Reznik09, Hahn09} found no measurable changes in most optic phonons across the magnetic transition, in contrast to what has been predicted by DFT calculations making allowance for spin-phonon coupling.  Therefore, the importance of spin-phonon coupling for the occurrence of high-\Tc\ superconductivity in the pnictides remains an open question.

The structural transition is accompanied by a transition into an antiferromagnetically ordered state. DFT calculations predict that the AFM state is energetically more stable even in the absence of a structural distortion, and that the structural distortion is the consequence of AFM order via magnetoelastic coupling.  The structural transition temperature (T$_s$) always occurs coincident with or at a higher temperature that the magnetic transition (T$_N$) in the electron-doped pnictides, which has led to the proposition of a nematic phase which does not show static magnetic order, in which magnetic fluctuations have a preferred direction \cite{Fernandes2010}. Alternatively, the nematic phase might be characterized by orbital ordering with unequal occupations of the d$_{xz}$ and d$_{yz}$ orbitals \cite{CCLee2009,Yi2011}.  The magnetic, orbital, and structural order parameters break the tetragonal symmetry in the same way, but the parameter primarily responsible for the symmetry breaking is still a matter of debate (e.g., recent work \cite{Liang2013} has claimed that the nematic state is driven by spin, but strongly enhanced by orbital fluctuations).  However as pointed out in Ref. \onlinecite{Fernandes2014}, if the parameters are strongly coupled the ``primary'' parameter loses some distinction.

In this paper, we investigate the coupling of the crystal lattice to the magnetic degrees of freedom by a study of transverse acoustic phonons and the magnetic properties of undoped \SFA\ and \BFA. More precisely, phonons were studied whose eigenvectors correspond to the structural distortion in the long-wavelength limit. These phonons propagate along the tetragonal \hkl[100] direction and are polarized along \hkl[010]. An early IXS study \cite{Niedziela2011} showed a pronounced softening of such phonons at low \qvec\ (reduced momentum transfer) on approaching on approaching \Ts\ from above. On further cooling, these modes hardened gradually to some extent. The softening on cooling is corroborated by measurements of the elastic modulus by ultrasound techniques \cite{Fernandes2010}. The softening was found to be closely correlated with a previous report \cite{Fernandes2013} of magnetic susceptibility as deduced from NMR above \TN. 

The motivation for our present study was twofold.  In the first place, whereas the behavior of the shear modulus\cite{Fernandes2010} and phonon frequencies above \TN\ reported so far looks quite plausible, this is not really the case for the behavior below \TN: since the phase transition in undoped pnictides is largely first-order, one would expect strong and sudden changes on cooling below \TN/\Ts.  In the second place, to gain a better understanding of the phonon behavior below the N\'eel temperature (the uncertainties associated with the lowest-temperature points in the previous report \cite{Niedziela2011} were large).  We used inelastic neutron scattering for our study which allowed more precise measurement of phonon energies over a wide range of temperatures.  Neutron scattering was further used to explore the magnetic properties of our samples above and below \TN\ with the aim to correlate the magnetic properties with the phonon properties.  We show that the magnetic correlation length above \TN\ and the magnetic order parameter squared below \TN\ track the phonon softening at all temperatures.

%%%%%%%%%%%%%%%%%%%%%%%%%%%%%%%%%%%%%%%%%%%%%%%%%%%%%%%%%%%%%%%%%%%%%%%%%%%%%%%%
\section{Experimental details}

We investigated two single-crystal samples \cite{Hardy2010} of \BFA~(Ba122) and of \SFA~(Sr122), each of mass~\appr\ 1~g, and mosaic spread of \textless~\degree{0.5}. The samples were mounted in closed-cycle refrigerators with a temperature range of 10~K to 300~K (\BFA) or 10~K to 350~K (\SFA). The experiments were carried out on the 4F cold triple-axis spectrometer and on the 1T thermal triple-axis spectrometer at the ORPH\`EE reactor at the Laboratoire L\'eon Brillouin at Saclay, France. On 1T, final energies were chosen between 8~meV and 14.7~meV, depending on the energy resolution needed for the particular part of the experiments.  The open collimations of the standard configuration were used for some scans (effectively 35'-35'-35'-35'), but tighter collimations were also often used to improve the resolution in energy and momentum transfer.  On 4F, measurements were made with initial energy of 8~meV, and used 50' in-pile collimation, 50' collimation between the pair of monochromators, and 40' collimation elsewhere.  Resolutions of 0.4~meV and 0.05~\AA$^{-1}$ were easily achieved.  Measurements of the phonon branch were made in the tetragonal HK0 plane. Most measurements of the magnetic fluctuations were made in the tetragonal HHL plane, but for some measurements we mounted the sample in the HK0 plane and tilted the sample to reach finite values of L.  Peaks were fit to Gaussian functions; error bars are derived from the statistical uncertainty.  Follow-up measurements were conducted using the BT-7 instrument at the NIST Center for Neutron Research.

%%%%%%%%%%
% 1
%%%%%%%%%%%
\begin{figure}[h]
   \centering
   \includegraphics[scale=0.22]{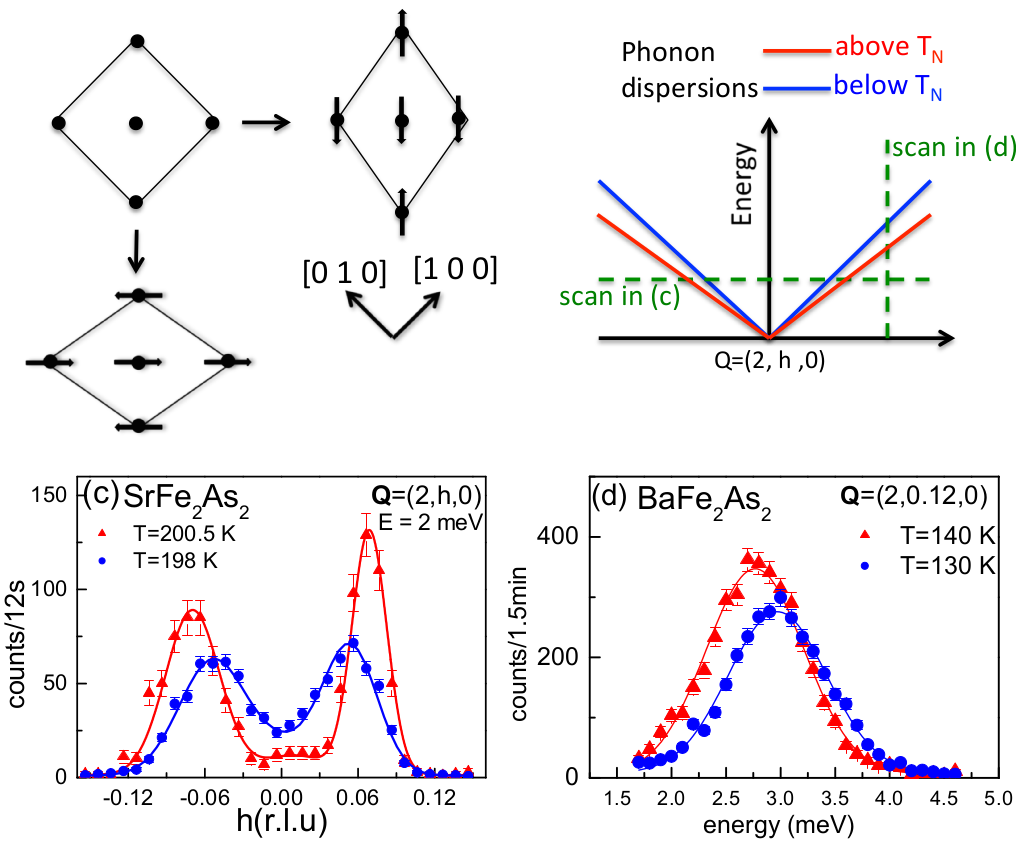} 
    \caption{(color online) Lattice deformation and phonon renormalization across the magnetic ordering transition. a) Schematic of spin-phonon coupling in iron pnictides. Dots represent Fe positions. Straight lines/arrows on top are boundaries of the unit cell/unit vectors used for the notation in this paper respectively. Parallel/antiparallel alignment of the spins on near-neighbor Fe sites favor longer/shorter bonds. b) Schematic of scans shown in (c) and (d). c) Constant-energy scans of \SFA\ at 2 meV showing that the acoustic phonon peaks move closer to the zone center upon cooling below \Ts~= 200~K, which indicates that the low-energy dispersion becomes steeper (schematic in (b), lines are a guide to the eye). d) constant-Q scans reflecting phonon hardening on cooling through \TN~= 135~K in \BFA (schematic in (b), lines are a guide to the eye).  All error bars represent one standard deviation, and are derived from the statistical uncertainty.
}
   \label{fig:Fig1}
\end{figure}
%%%%%%%%%%%%

%%%%%%%%%%%%%%%%%%%%%%%%%%%%%%%%%%%%%%%%%%%%%%%%%%%%%%%%%%%%%%%%%%%%%%%%%%%%%%%%
\section{Results}

%%%%%%%%%%
% 2
%%%%%%%%%%%
\begin{figure*}[t]
   \centering
   \includegraphics[scale=0.55]{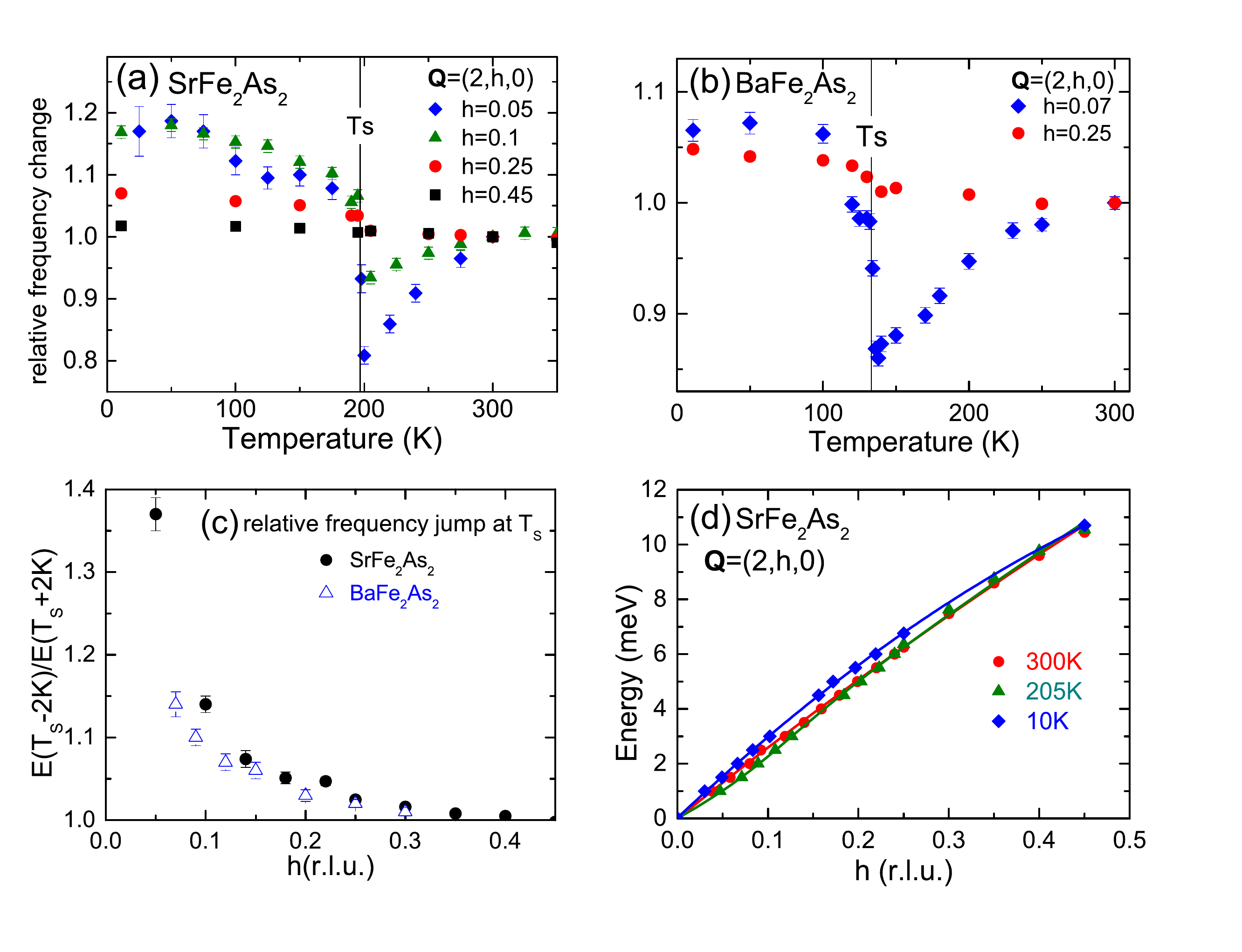} 
    \caption{(color online) Phonon renormalization in \SFA\ and \BFA\ as a function of temperature along \hkl[100]. Error bars are drawn in all panels, but are sometimes smaller than the symbol.  a,b) relative phonon frequency change as a function of temperature in \SFA\ (a) and \BFA\ (b) c) Relative frequency jump at \Ts. d) Dispersion of transverse phonons propagating along \hkl[100] with polarization along \hkl[010] in \SFA.  Lines are a guide to the eye.  All error bars represent one standard deviation, and are derived from the statistical uncertainty.
}
   \label{fig:Fig2}
\end{figure*}
%%%%%%%%%%%%

At room temperature these materials are tetragonal, but as has been amply documented in the literature \cite{Tegel08-1, Rotter08}, they undergo a structural distortion from tetragonal-to-orthorhombic structure at \Ts.  This transition is closely accompanied by a transition to an antiferromagnetically ordered state at \TN. \TN\ is very close to \Ts, if different at all.  We did not observe any separation of the structural and AFM transitions in either of our samples.  The transition was observed at T~= 200.0(5)~K (\SFA) and 135(1)~K (\BFA). The transition can be understood within the framework of DFT calculations, which show that magnetic order as observed in experiment with antiparallel spin alignment on neighboring Fe sites in one direction and parallel alignment in the other direction leads to a lowering of the total energy \cite{Yildirim09-1}. A further energy gain is obtained by shortening the bonds with parallel spin alignment and stretching the bonds with antiparallel spin alignment, as illustrated in Fig.~\ref{fig:Fig1}a. The transition into the orthorhombic state leads to twinning of the samples, which results in a noticeable broadening of the mosaic distribution. In the following, we will adopt the tetragonal notation, and will make use of the orthorhombic notation only in a few places.

%%%%%%%%%%
% 3
%%%%%%%%%%%
\begin{figure*}[t]
   \centering
   \includegraphics[scale=0.55]{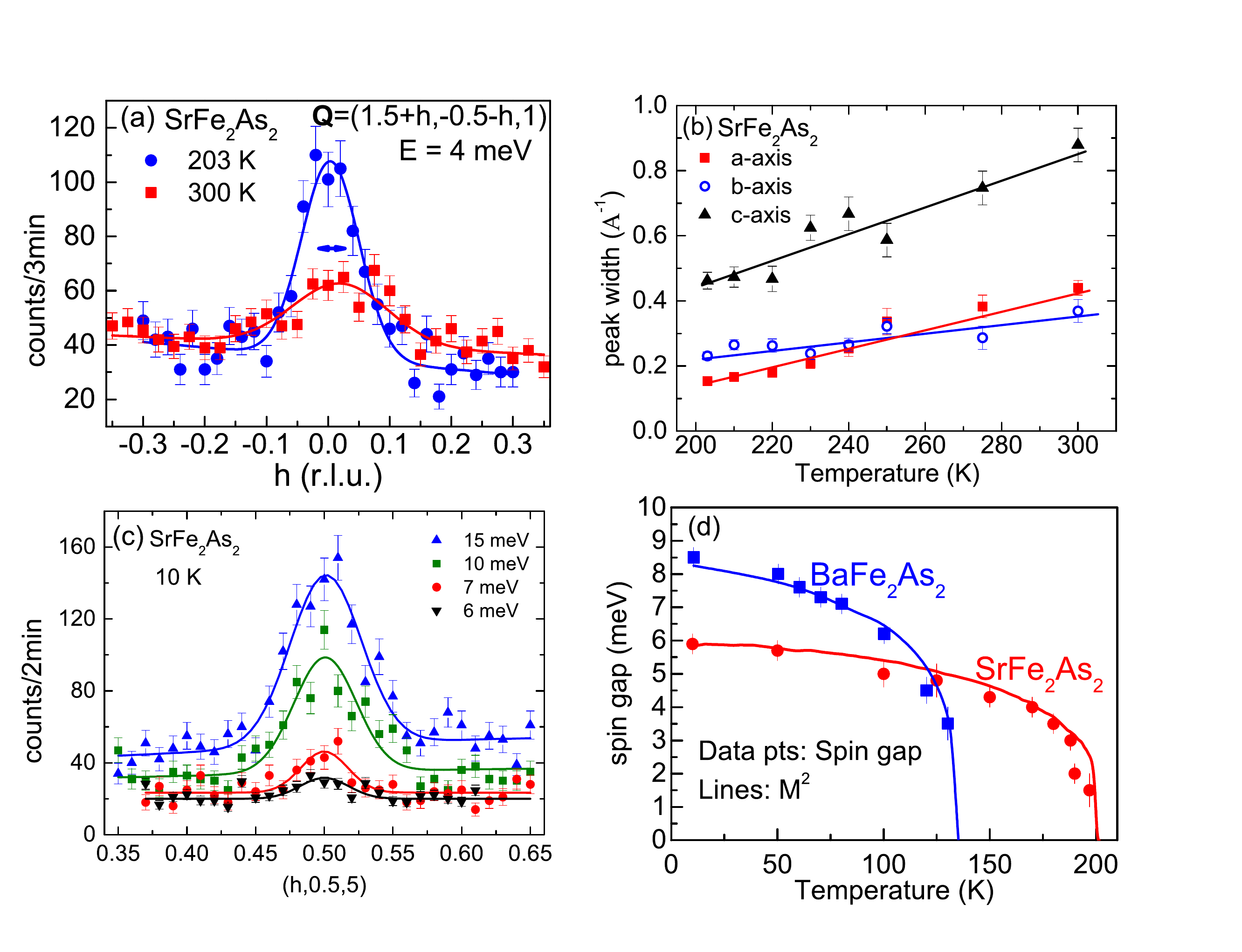} 
    \caption{(color online) Magnetic fluctuations in \SFA\ and \BFA. a) Constant-energy (E~= 4~meV) scans through the magnetic superlattice peak at two temperatures. Lines are fits with a Gaussian. The horizontal blue arrow denotes the experimental resolution.  b) Resolution-corrected widths of spin fluctuation peaks measured at E~= 4~meV versus temperature.  The labeling corresponds to the orthorhombic axes.  c) Constant-energy scans from which the spin gap was determined d) Spin gap and the square of the magnetic order parameter versus temperature. M$^{2}$ has been scaled to match the value of the spin gap.  All error bars represent one standard deviation, and are derived from the statistical uncertainty.
}
   \label{fig:Fig3}
\end{figure*}
%%%%%%%%%%%%

As discussed in the introduction, the long-wavelength transverse acoustic phonons are expected to soften on approaching \Ts~from above, followed by a gradual and moderate hardening on cooling below \Ts.  To our surprise, we observed a strong and sudden hardening of this branch in \SFA\ (typically phonon energies might change by 1\% over a temperature change of 100~K; the energy of the low-\qvec\ portion of this branch changes by over 20\% within just a few degrees).  The previous investigation made by IXS was, however, performed on \BFA.  In order to see whether the two compounds behave in a different way, or whether the effect observed in \SFA\ is a generic feature of the ferropnictides, complementary measurements were performed on \BFA.  It turned out that the two compounds show very similar behavior.  The phonon effect is clearly evident from both constant-\Qvec\ (constant momentum transfer) and constant-energy scans through the phonon dispersions (see Fig.~\ref{fig:Fig1}c,d).  The constant-\Qvec\ scan in Fig.~\ref{fig:Fig1}d illustrates that in \BFA\ the phonon hardens strongly on cooling across \TN~= 135~K by just 10~K from 140~K to 130~K.  The same effect is reflected in the constant-energy scan on Sr122 in Fig.~\ref{fig:Fig1}c.  Here \qvec~=~0 is the zone center and the peaks on both sides at \qvec~=~0.07 r.l.u. (reciprocal lattice units) originate from acoustic phonons dispersing on both sides of the zone center. The steeper (shallower) the dispersions, the closer (further) the peaks in the constant-energy scan are to the zone center (see schematic in Fig.~\ref{fig:Fig1}b).  On cooling through \TN~= 200~K, the peaks move closer together, which reflects the same hardening as observed in Fig.~\ref{fig:Fig1}c.  There is some broadening in both \Qvec~and energy (some part of which is certainly due to the broader mosaic arising from twinning in the orthorhombic phase), but the change in energy is nonetheless clear.

Fig.~\ref{fig:Fig2}a shows varying behavior at different \qvec.  For small \qvec, the phonon softens substantially upon cooling towards \TN, which is followed by abrupt hardening at lower temperature. At intermediate \qvec~(e.g. \qvec~=~0.25) there is no softening above \TN, but there is still the hardening below \TN. And at high \qvec~(e.g. \qvec~=~0.45), no temperature dependence is observed across \TN. The ratio of the frequency just below \TN\ divided by the frequency above \TN\ increases strongly towards the zone center (Fig.~\ref{fig:Fig2}a,b). As is evident from Fig.~\ref{fig:Fig2}c, the hardening below \TN\ not only undoes the softening above \TN, but overshoots it by a substantial amount (see Fig.~\ref{fig:Fig2}d). Further, the hardening below \TN\ extends much farther in \qvec, pointing to a different origin of the two phenomena.

Since we expected the phonon effect to be related to the formation of magnetic order, we carefully investigated the temperature dependence of magnetic Bragg peak intensity below \TN\ and of magnetic fluctuations above \TN\ in the same samples.  In order to correlate the phonon softening above \TN\ with magnetic properties, we measured the width in momentum transfer of the magnetic fluctuations at several low energies (up to 6~meV).  The \qvec-resolution at each energy was determined empirically from the spin wave spectrum below \TN\ (which is nearly resolution-limited at the energies considered).  We found that the temperature dependence of the linewidths was essentially the same at all energies investigated.  The data shown are those for which we obtained the best quality.  Fig.~\ref{fig:Fig3}a shows that well-defined peaks from magnetic scattering at 4~meV are already present at 300~K. They become much sharper on approaching \TN\ from above, although they never become resolution-limited, even close to \TN.  As generally expected, the in-plane widths are considerably smaller than those measured along c, but even in the c-direction, linewidth narrowing begins already at 300~K (Fig.~\ref{fig:Fig3}b). This highlights the three-dimensional character of low energy magnetic fluctuations, and is in qualitative agreement with previous reports on the spin dynamics in the \BFA\ and \SFA\ systems \cite{Ewings2011,Matan10}. None of the widths taken alone correlates well with the phonon softening above \TN, however, the product of all three does (Fig. ~\ref{fig:Fig4}b). Since each linewidth is proportional to the inverse of a correlation length in a particular direction, the product of the linewidths is proportional to the inverse of the volume of the correlated domains.

Below \TN, we determined both the temperature dependence of the magnetic order parameter and that of the spin gap. For the magnetic order parameter, we measured the intensity of the magnetic Bragg peak at \Qvec~= \hkl(2.5, 2.5, 1) (or \hkl(5 0 1) in orthorhombic notation). To determine the spin gap, we made a series of constant-energy scans through the magnetic Bragg peak at \Qvec~= \hkl(0.5,~0.5,~5) (or \hkl(105) in orthorhombic notation, see Fig.~\ref{fig:Fig3}a). On approaching the spin gap energy from above, the peak intensities decrease sharply for reasons of phase space.  We made careful measurements of the spin gap energy, and found that the temperature dependence of the spin gap tracks the square of the magnetic moment quite closely, for both Sr122 and Ba122 (see Fig.~\ref{fig:Fig3}d).

%%%%%%%%%%%%%%%%%%%%%%%%%%%%%%%%%%%%%%%%%%%%%%%%%%%%%%%%%%%%%%%%%%%%%%%%%%%%%%%%
\section{Discussion}

%%%%%%%%%%
% 4
%%%%%%%%%%%
\begin{figure}[h!]
   \centering
   \includegraphics[scale=0.45]{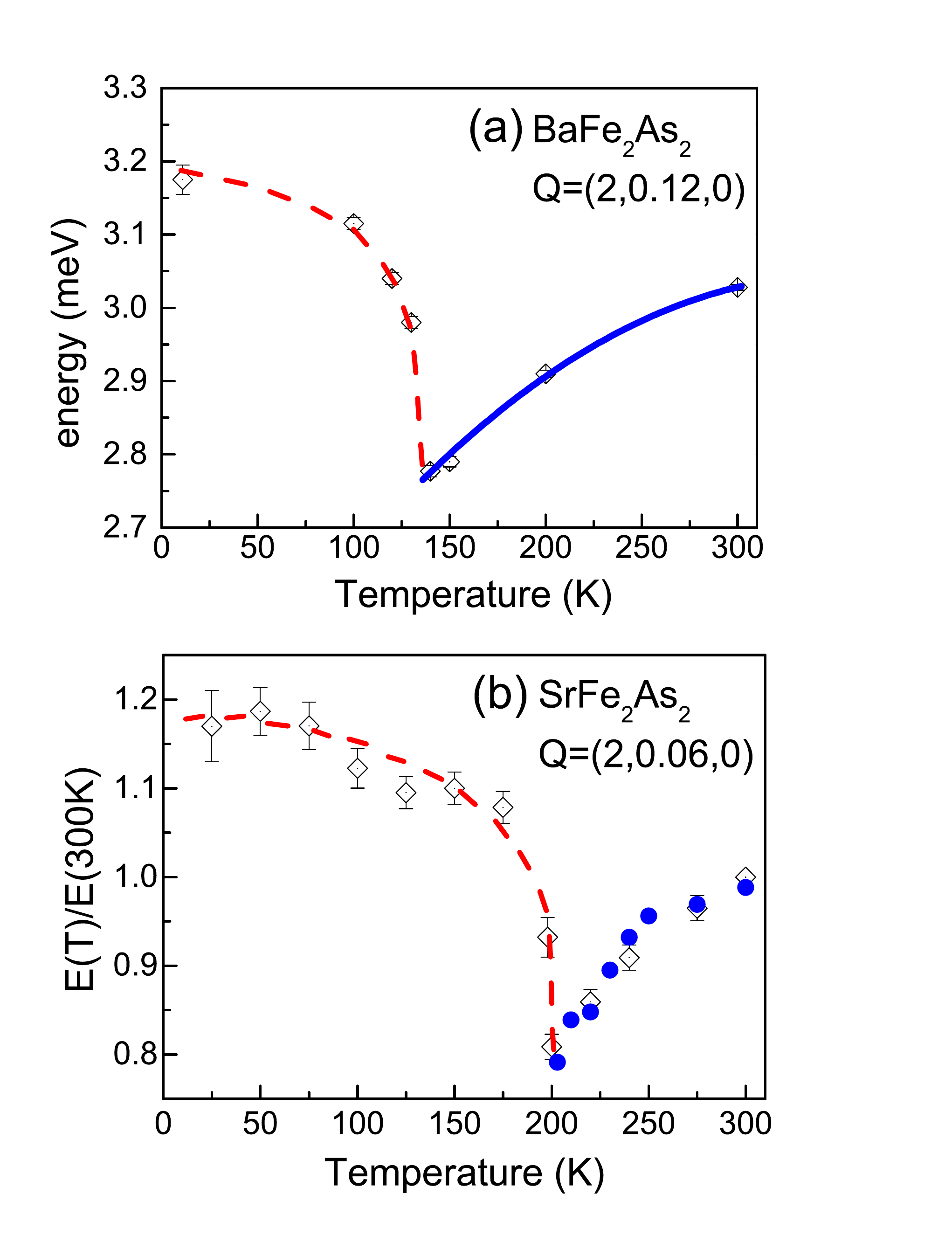} 
    \caption{(color online) Correspondence between the renormalized phonon energies and magnetic properties in \BFA\ and \SFA.  Open markers represent phonon frequencies, and red dashed lines represent peak intensities of the magnetic Bragg peaks.  a) For \BFA, the solid blue line is a guide to the eye.  b) for \SFA, the filled blue circles represent 1/$\xi$ (scaled to the phonon data), where $\xi$ is the size of slowly-fluctuating magnetic domains, determined by multiplying inverses of all three linewidths shown in Fig.~\ref{fig:Fig3}b.  All error bars represent one standard deviation, and are derived from the statistical uncertainty.
}
   \label{fig:Fig4}
\end{figure}
%%%%%%%%%%%%

%T_below:
%	q=0
%	q_low
%	q_high

For temperatures below \TN, it is clear that our data show a very large hardening that exceeds the softening above \TN. In the \qvec~=~0 limit, these results should extrapolate to the shear modulus that was previously reported for \BFA\ based on resonant ultrasound experiments (RUS)\cite{Fernandes2010}. However, RUS shows only a slight hardening on cooling (small in comparison to the softening above \TN).  This disagreement is likely due to the twin domain boundaries inside the crystal in the orthorhombic phase, which are known to interfere with RUS measurement, but not with phonon measurements (beyond a slight broadening).  In addition, there is a significant quantitative disagreement between our data and the previous IXS report \cite{Niedziela2011} for T~$<$~\TN, in that the low-\qvec\ IXS results for \qvec~$\leq$~0.1 do not capture the sudden and very strong increase of phonon frequencies just below \TN.  We do not have any plausible explanation for the disagreement, but emphasize that we observe the strong and sudden increase of phonon frequencies below \TN\ under a wide range of experimental conditions (multiple samples, growers, instruments, sample environments, etc.).

We found that below \TN, the phonon hardening tracks the square of the magnetic order parameter very closely in both \SFA\ and in \BFA\ (Fig.~\ref{fig:Fig4}) (although the effect is somewhat smaller in \BFA, possibly due to the greater mass).  Since we have found that the low-\qvec\ phonon frequency is biquadratically coupled to the magnetic order parameter, it is essentially linearly coupled to the structural order parameter (in the undoped 122-compounds the magnetic and the structural order parameters are biquadratically coupled \cite{Avci2011, Cano2010}).  But since we also found that the temperature evolution of the spin gap is quite similar to that of the square of the magnetic order parameter, the correlation seen in Fig.~\ref{fig:Fig3}d might instead be related to the opening of the spin gap.  There thus seem to be three possibilities for the correlation observed in Fig.~\ref{fig:Fig4}: that the phonon hardening could track either the square of the structural distortion, the square of the magnetic order parameter, or opening of the spin gap.  To determine which is this case, further studies are necessary on samples where the structural and the magnetic order parameters do not follow the same temperature dependence, as in, e.g., Co-doped Ba122.

%T_above:
%	q=0
%	q_low
%	q_high

At temperatures above \TN\ the softening of the investigated phonon modes is proportional to the volume of fluctuating magnetic domains.  Our data are qualitatively consistent with the RUS results above \TN: the slope at the lowest \qvec\ becomes rapidly softer upon approaching \TN from above.  A direct quantitative comparison is not possible, because the softening is strongly dependent on \qvec, and imperfect \qvec-resolution does not allow us to probe \qvec-values which are very close to \qvec~=~0.

When considering finite values of \qvec, we find that for \qvec~$\geq$  0.2~r.l.u., the energy of the phonon branch is nearly unaffected when approaching \TN\ from above.  For \qvec~$\leq$ 0.2~r.l.u., the branch softens rapidly upon approaching \TN.  The absolute value of the softening is greatest close to \qvec~= 0.1~r.l.u., and the relative value of the softening is greatest at the lowest measured \qvec\ (\appr~ 0.05~r.l.u.).  These results are also fully consistent with the IXS results for \BFA\ above \TN\ reported in Ref. \onlinecite{Niedziela2011}, after taking into account the somewhat coarser \qvec-resolution used here.

We also find that above \TN\, there is a close correlation between the volume of the short-lived magnetic domains and the phonon frequencies (Fig.~\ref{fig:Fig4}).  As previously shown \cite{ChuJH2010}, only a slight uniaxial strain is required to produce a resistivity anisotropy in the paramagnetic state, which corresponds to the soft phonon seen here.  Since a local lattice distortion will entail an elastic strain field, which is inherently long-range in 3D, it is plausible that the volume of the dynamic domains is closely correlated to the phonon frequencies.  It follows that in order to understand coupling of magnetic (and for that matter nematic) fluctuations to the atomic lattice, it is necessary to consider not only the in-plane magnetic correlation length but also the correlation length along the c-axis.

Although it is clear that there exists a strong coupling between the lattice and the electronic degrees of freedom, it can be difficult to distinguish the primary order parameter (given that the magnetic, orbital, and structural order parameters break the tetragonal symmetry in the same way).  In this context it may be illuminating to consider the \qvec-dependence of phonon softening, as its correlation with the \qvec-dependence of magnetic or orbital fluctuations make it possible to differentiate between the two mechanisms. This question requires a theoretical investigation that is outside the scope of the current work.  In any event, our finite-\qvec\ data provide detailed information relating a microscopic quantity (the size of the magnetic domains) to a macroscopic quantity (the elastic constant), and may prove to be a useful basis for comparison with theory.

%% further discussion
It seems likely that there are two competing effects; while the softening above \TN\ is primarily confined to \qvec~$\leq$~0.1, the hardening extends to much higher \qvec.  Both effects are associated with the \hkl(110) transverse shear mode.  We took measurements of the longitudinal acoustic phonon along the \hkl(110) using the BT-7 instrument (data not shown) and found no evidence of either effect there, nor did we see any strong changes of the high-energy phonons in our previous inelastic neutron scattering study\cite{Parshall2014} of \SFA.

The previous IXS study suggested the softening might occur as a result of a Kohn anomaly due to two-dimensional Fermi surface nesting, because the  range of the softening corresponds roughly to the size of inner hole pocket in \BFA\ (\appr~0.1~r.l.u.).  In that case, there should be an enhancement of the electron-phonon coupling in the AFM state (previous theoretical work found that the electron-phonon coupling was significantly stronger in the stripe AFM state than in the paramagnetic state \cite{Boeri2010}).  It is difficult for us to determine if the electron-phonon coupling for this mode (as measured by the energy linewidth) has become significantly enhanced in the AFM phase.  This is because the orthorhombic distortion causes a broadening of the mosaic spread, which in turn causes a broadening of the TA phonon.  Separating these effects is challenging.

It may be possible to understand the detailed \qvec-dependence of the hardening using DFT calculations.  In order to capture both the low-\qvec\ deviation from linear dispersion, and then the gradual return to the paramagnetic behavior at high-\qvec, one would need to calculate a very dense mesh in \qvec\, for both the AFM and paramagnetic phases.  Such a calculation would be time-consuming, and given the challenges associated with DFT in this system \cite{Mazin08-1}, we would have only limited faith in the results.

%%%%%%%%%%%%%%%%%%%%%%%%%%%%%%%%%%%%%%%%%%%%%%%%%%%%%%%%%%%%%%%%%%%%%%%%%%%%%%%%
\section{Conclusions}

We found that in undoped 122 compounds, the TA phonons propagating along \hkl[100] with polarization along \hkl[010] soften gradually on approaching \TN\ from above, followed by an abrupt hardening just below \TN\ and further gradual hardening on cooling to very low temperatures. Interestingly, the hardening below \TN\ exceeds significantly that of the softening above \TN\ in both magnitude and \qvec-range, which remains to be understood. While the softening above \TN\ was known already from previous publications at least qualitatively, the pronounced hardening below \TN\ was not. The temperature evolution of the phonon frequencies both below and above \TN\ correlates closely with that of magnetic properties observed on the same samples. However, this correlation does not necessarily imply a direct spin-phonon coupling. The coupling between the phonons and the static/dynamic magnetic order might only be indirect via magnetoelastic coupling.

The most promising candidates for further insight into this issue are samples where magnetic order and structural distortions do not appear at the same temperature, as in Ba122 samples lightly doped with Co. In such samples, \Ts~$>$~\TN. Just like in the undoped sample, we can expect gradual softening of the phonons on cooling towards the structural transition. It is, however, difficult to predict how the phonon frequencies will evolve on further cooling through the magnetic transition. Answering this question will be a subject of a different paper, which will provide additional insight into understanding the coupling of phonons to magnetic properties in Fe-based superconductors.

%%%%%%%%%%%%%%%%%%%%%%%%%%%%%%%%%%%%%%%%%%%%%%%%%%%%%%%%%%%%%%%%%%%%%%%%%%%%%%%%
% >>> Acknowledgments <<<
\begin{acknowledgments}
The research at ORNL's Spallation Neutron Source was sponsored by the Scientific User Facilities Division, Office of Basic Energy Sciences, U.S. Department of Energy.  D.P. and D.R. were supported by the U.S. Department of Energy, Office of Basic Energy Sciences, Office of Science, under Contract No. DE-SC0006939. The authors thank A. Alatas for valuable discussions.
\end{acknowledgments}

%%%%%%%%%%%%%%%%%%%%%%%%%%%%%%%%%%%%%%%%%%%%%%%%%%%%%%%%%%%%%%%%%%%%%%%%%%%%%%%%
% >>> Bibliography <<<
%\bibliography{JabRef_bibliography}

\begin{thebibliography}{27}%
\makeatletter
\providecommand \@ifxundefined [1]{%
 \@ifx{#1\undefined}
}%
\providecommand \@ifnum [1]{%
 \ifnum #1\expandafter \@firstoftwo
 \else \expandafter \@secondoftwo
 \fi
}%
\providecommand \@ifx [1]{%
 \ifx #1\expandafter \@firstoftwo
 \else \expandafter \@secondoftwo
 \fi
}%
\providecommand \natexlab [1]{#1}%
\providecommand \enquote  [1]{``#1''}%
\providecommand \bibnamefont  [1]{#1}%
\providecommand \bibfnamefont [1]{#1}%
\providecommand \citenamefont [1]{#1}%
\providecommand \href@noop [0]{\@secondoftwo}%
\providecommand \href [0]{\begingroup \@sanitize@url \@href}%
\providecommand \@href[1]{\@@startlink{#1}\@@href}%
\providecommand \@@href[1]{\endgroup#1\@@endlink}%
\providecommand \@sanitize@url [0]{\catcode `\\12\catcode `\$12\catcode
  `\&12\catcode `\#12\catcode `\^12\catcode `\_12\catcode `\%12\relax}%
\providecommand \@@startlink[1]{}%
\providecommand \@@endlink[0]{}%
\providecommand \url  [0]{\begingroup\@sanitize@url \@url }%
\providecommand \@url [1]{\endgroup\@href {#1}{\urlprefix }}%
\providecommand \urlprefix  [0]{URL }%
\providecommand \Eprint [0]{\href }%
\providecommand \doibase [0]{http://dx.doi.org/}%
\providecommand \selectlanguage [0]{\@gobble}%
\providecommand \bibinfo  [0]{\@secondoftwo}%
\providecommand \bibfield  [0]{\@secondoftwo}%
\providecommand \translation [1]{[#1]}%
\providecommand \BibitemOpen [0]{}%
\providecommand \bibitemStop [0]{}%
\providecommand \bibitemNoStop [0]{.\EOS\space}%
\providecommand \EOS [0]{\spacefactor3000\relax}%
\providecommand \BibitemShut  [1]{\csname bibitem#1\endcsname}%
\let\auto@bib@innerbib\@empty
%</preamble>
\bibitem [{\citenamefont {Dai}\ \emph {et~al.}(2012)\citenamefont {Dai},
  \citenamefont {Hu},\ and\ \citenamefont {Dagotto}}]{Dai2012}%
  \BibitemOpen
  \bibfield  {author} {\bibinfo {author} {\bibfnamefont {P.}~\bibnamefont
  {Dai}}, \bibinfo {author} {\bibfnamefont {J.}~\bibnamefont {Hu}}, \ and\
  \bibinfo {author} {\bibfnamefont {E.}~\bibnamefont {Dagotto}},\ }\href
  {\doibase 10.1038/nphys2438} {\bibfield  {journal} {\bibinfo  {journal}
  {Nature Physics}\ }\textbf {\bibinfo {volume} {8}},\ \bibinfo {pages} {709}
  (\bibinfo {year} {2012})}\BibitemShut {NoStop}%
\bibitem [{\citenamefont {Yildirim}(2009)}]{Yildirim09-1}%
  \BibitemOpen
  \bibfield  {author} {\bibinfo {author} {\bibfnamefont {T.}~\bibnamefont
  {Yildirim}},\ }\href {\doibase 10.1016/j.physc.2009.03.038} {\bibfield
  {journal} {\bibinfo  {journal} {Physica C: Superconductivity}\ }\textbf
  {\bibinfo {volume} {469}},\ \bibinfo {pages} {425} (\bibinfo {year}
  {2009})}\BibitemShut {NoStop}%
\bibitem [{\citenamefont {Kim}\ \emph {et~al.}(2012)\citenamefont {Kim},
  \citenamefont {Pashkin}, \citenamefont {Schäfer}, \citenamefont {Beyer},
  \citenamefont {Porer}, \citenamefont {Wolf}, \citenamefont {Bernhard},
  \citenamefont {Demsar}, \citenamefont {Huber},\ and\ \citenamefont
  {Leitenstorfer}}]{KimKW2012}%
  \BibitemOpen
  \bibfield  {author} {\bibinfo {author} {\bibfnamefont {K.~W.}\ \bibnamefont
  {Kim}}, \bibinfo {author} {\bibfnamefont {A.}~\bibnamefont {Pashkin}},
  \bibinfo {author} {\bibfnamefont {H.}~\bibnamefont {Schäfer}}, \bibinfo
  {author} {\bibfnamefont {M.}~\bibnamefont {Beyer}}, \bibinfo {author}
  {\bibfnamefont {M.}~\bibnamefont {Porer}}, \bibinfo {author} {\bibfnamefont
  {T.}~\bibnamefont {Wolf}}, \bibinfo {author} {\bibfnamefont {C.}~\bibnamefont
  {Bernhard}}, \bibinfo {author} {\bibfnamefont {J.}~\bibnamefont {Demsar}},
  \bibinfo {author} {\bibfnamefont {R.}~\bibnamefont {Huber}}, \ and\ \bibinfo
  {author} {\bibfnamefont {A.}~\bibnamefont {Leitenstorfer}},\ }\href {\doibase
  10.1038/nmat3294} {\bibfield  {journal} {\bibinfo  {journal} {Nature
  Materials}\ }\textbf {\bibinfo {volume} {11}},\ \bibinfo {pages} {497}
  (\bibinfo {year} {2012})}\BibitemShut {NoStop}%
\bibitem [{\citenamefont {Chauvi\`ere}\ \emph {et~al.}(2009)\citenamefont
  {Chauvi\`ere}, \citenamefont {Gallais}, \citenamefont {Cazayous},
  \citenamefont {Sacuto}, \citenamefont {M\'easson}, \citenamefont {Colson},\
  and\ \citenamefont {Forget}}]{Chauviere2009}%
  \BibitemOpen
  \bibfield  {author} {\bibinfo {author} {\bibfnamefont {L.}~\bibnamefont
  {Chauvi\`ere}}, \bibinfo {author} {\bibfnamefont {Y.}~\bibnamefont
  {Gallais}}, \bibinfo {author} {\bibfnamefont {M.}~\bibnamefont {Cazayous}},
  \bibinfo {author} {\bibfnamefont {A.}~\bibnamefont {Sacuto}}, \bibinfo
  {author} {\bibfnamefont {M.~A.}\ \bibnamefont {M\'easson}}, \bibinfo {author}
  {\bibfnamefont {D.}~\bibnamefont {Colson}}, \ and\ \bibinfo {author}
  {\bibfnamefont {A.}~\bibnamefont {Forget}},\ }\href {\doibase
  10.1103/PhysRevB.80.094504} {\bibfield  {journal} {\bibinfo  {journal} {Phys.
  Rev. B}\ }\textbf {\bibinfo {volume} {80}},\ \bibinfo {pages} {094504}
  (\bibinfo {year} {2009})}\BibitemShut {NoStop}%
\bibitem [{\citenamefont {Egami}\ \emph {et~al.}(2010)\citenamefont {Egami},
  \citenamefont {Fine}, \citenamefont {Singh}, \citenamefont {Parshall},
  \citenamefont {de~la Cruz},\ and\ \citenamefont {Dai}}]{Egami10-1}%
  \BibitemOpen
  \bibfield  {author} {\bibinfo {author} {\bibfnamefont {T.}~\bibnamefont
  {Egami}}, \bibinfo {author} {\bibfnamefont {B.}~\bibnamefont {Fine}},
  \bibinfo {author} {\bibfnamefont {D.}~\bibnamefont {Singh}}, \bibinfo
  {author} {\bibfnamefont {D.}~\bibnamefont {Parshall}}, \bibinfo {author}
  {\bibfnamefont {C.}~\bibnamefont {de~la Cruz}}, \ and\ \bibinfo {author}
  {\bibfnamefont {P.}~\bibnamefont {Dai}},\ }\href {\doibase
  10.1016/j.physc.2009.11.167} {\bibfield  {journal} {\bibinfo  {journal}
  {Physica C: Superconductivity}\ }\textbf {\bibinfo {volume} {470}},\ \bibinfo
  {pages} {S294} (\bibinfo {year} {2010})}\BibitemShut {NoStop}%
\bibitem [{\citenamefont {Yndurain}(2011)}]{Yndurain2011}%
  \BibitemOpen
  \bibfield  {author} {\bibinfo {author} {\bibfnamefont {F.}~\bibnamefont
  {Yndurain}},\ }\href@noop {} {\bibfield  {journal} {\bibinfo  {journal}
  {Europhysics Letters (EPL)}\ }\textbf {\bibinfo {volume} {94}},\ \bibinfo
  {pages} {37001} (\bibinfo {year} {2011})}\BibitemShut {NoStop}%
\bibitem [{\citenamefont {Boeri}\ \emph {et~al.}(2009)\citenamefont {Boeri},
  \citenamefont {Dolgov},\ and\ \citenamefont {Golubov}}]{Boeri09}%
  \BibitemOpen
  \bibfield  {author} {\bibinfo {author} {\bibfnamefont {L.}~\bibnamefont
  {Boeri}}, \bibinfo {author} {\bibfnamefont {O.}~\bibnamefont {Dolgov}}, \
  and\ \bibinfo {author} {\bibfnamefont {A.}~\bibnamefont {Golubov}},\ }\href
  {\doibase 10.1016/j.physc.2009.03.020} {\bibfield  {journal} {\bibinfo
  {journal} {Physica C: Superconductivity}\ }\textbf {\bibinfo {volume}
  {469}},\ \bibinfo {pages} {628} (\bibinfo {year} {2009})}\BibitemShut
  {NoStop}%
\bibitem [{\citenamefont {Reznik}\ \emph {et~al.}(2009)\citenamefont {Reznik},
  \citenamefont {Lokshin}, \citenamefont {Mitchell}, \citenamefont {Parshall},
  \citenamefont {Dmowski}, \citenamefont {Lamago}, \citenamefont {Heid},
  \citenamefont {Bohnen}, \citenamefont {Sefat}, \citenamefont {McGuire},
  \citenamefont {Sales}, \citenamefont {Mandrus}, \citenamefont {Subedi},
  \citenamefont {Singh}, \citenamefont {Alatas}, \citenamefont {Upton},
  \citenamefont {Said}, \citenamefont {Cunsolo}, \citenamefont {Shvyd{'}ko},\
  and\ \citenamefont {Egami}}]{Reznik09}%
  \BibitemOpen
  \bibfield  {author} {\bibinfo {author} {\bibfnamefont {D.}~\bibnamefont
  {Reznik}}, \bibinfo {author} {\bibfnamefont {K.}~\bibnamefont {Lokshin}},
  \bibinfo {author} {\bibfnamefont {D.~C.}\ \bibnamefont {Mitchell}}, \bibinfo
  {author} {\bibfnamefont {D.}~\bibnamefont {Parshall}}, \bibinfo {author}
  {\bibfnamefont {W.}~\bibnamefont {Dmowski}}, \bibinfo {author} {\bibfnamefont
  {D.}~\bibnamefont {Lamago}}, \bibinfo {author} {\bibfnamefont
  {R.}~\bibnamefont {Heid}}, \bibinfo {author} {\bibfnamefont {K.-P.}\
  \bibnamefont {Bohnen}}, \bibinfo {author} {\bibfnamefont {A.~S.}\
  \bibnamefont {Sefat}}, \bibinfo {author} {\bibfnamefont {M.~A.}\ \bibnamefont
  {McGuire}}, \bibinfo {author} {\bibfnamefont {B.~C.}\ \bibnamefont {Sales}},
  \bibinfo {author} {\bibfnamefont {D.~G.}\ \bibnamefont {Mandrus}}, \bibinfo
  {author} {\bibfnamefont {A.}~\bibnamefont {Subedi}}, \bibinfo {author}
  {\bibfnamefont {D.~J.}\ \bibnamefont {Singh}}, \bibinfo {author}
  {\bibfnamefont {A.}~\bibnamefont {Alatas}}, \bibinfo {author} {\bibfnamefont
  {M.~H.}\ \bibnamefont {Upton}}, \bibinfo {author} {\bibfnamefont {A.~H.}\
  \bibnamefont {Said}}, \bibinfo {author} {\bibfnamefont {A.}~\bibnamefont
  {Cunsolo}}, \bibinfo {author} {\bibfnamefont {Y.}~\bibnamefont {Shvyd{'}ko}},
  \ and\ \bibinfo {author} {\bibfnamefont {T.}~\bibnamefont {Egami}},\ }\href
  {\doibase 10.1103/PhysRevB.80.214534} {\bibfield  {journal} {\bibinfo
  {journal} {Physical Review B}\ }\textbf {\bibinfo {volume} {80}},\ \bibinfo
  {pages} {214534} (\bibinfo {year} {2009})}\BibitemShut {NoStop}%
\bibitem [{\citenamefont {Hahn}\ \emph {et~al.}(2009)\citenamefont {Hahn},
  \citenamefont {Lee}, \citenamefont {Ni}, \citenamefont {Canfield},
  \citenamefont {Goldman}, \citenamefont {McQueeney}, \citenamefont {Harmon},
  \citenamefont {Alatas}, \citenamefont {Leu}, \citenamefont {Alp},
  \citenamefont {Chung}, \citenamefont {Todorov},\ and\ \citenamefont
  {Kanatzidis}}]{Hahn09}%
  \BibitemOpen
  \bibfield  {author} {\bibinfo {author} {\bibfnamefont {S.~E.}\ \bibnamefont
  {Hahn}}, \bibinfo {author} {\bibfnamefont {Y.}~\bibnamefont {Lee}}, \bibinfo
  {author} {\bibfnamefont {N.}~\bibnamefont {Ni}}, \bibinfo {author}
  {\bibfnamefont {P.~C.}\ \bibnamefont {Canfield}}, \bibinfo {author}
  {\bibfnamefont {A.~I.}\ \bibnamefont {Goldman}}, \bibinfo {author}
  {\bibfnamefont {R.~J.}\ \bibnamefont {McQueeney}}, \bibinfo {author}
  {\bibfnamefont {B.~N.}\ \bibnamefont {Harmon}}, \bibinfo {author}
  {\bibfnamefont {A.}~\bibnamefont {Alatas}}, \bibinfo {author} {\bibfnamefont
  {B.~M.}\ \bibnamefont {Leu}}, \bibinfo {author} {\bibfnamefont {E.~E.}\
  \bibnamefont {Alp}}, \bibinfo {author} {\bibfnamefont {D.~Y.}\ \bibnamefont
  {Chung}}, \bibinfo {author} {\bibfnamefont {I.~S.}\ \bibnamefont {Todorov}},
  \ and\ \bibinfo {author} {\bibfnamefont {M.~G.}\ \bibnamefont {Kanatzidis}},\
  }\href {\doibase 10.1103/PhysRevB.79.220511} {\bibfield  {journal} {\bibinfo
  {journal} {Phys. Rev. B}\ }\textbf {\bibinfo {volume} {79}},\ \bibinfo
  {pages} {220511} (\bibinfo {year} {2009})}\BibitemShut {NoStop}%
\bibitem [{\citenamefont {Fernandes}\ \emph {et~al.}(2010)\citenamefont
  {Fernandes}, \citenamefont {VanBebber}, \citenamefont {Bhattacharya},
  \citenamefont {Chandra}, \citenamefont {Keppens}, \citenamefont {Mandrus},
  \citenamefont {McGuire}, \citenamefont {Sales}, \citenamefont {Sefat},\ and\
  \citenamefont {Schmalian}}]{Fernandes2010}%
  \BibitemOpen
  \bibfield  {author} {\bibinfo {author} {\bibfnamefont {R.~M.}\ \bibnamefont
  {Fernandes}}, \bibinfo {author} {\bibfnamefont {L.~H.}\ \bibnamefont
  {VanBebber}}, \bibinfo {author} {\bibfnamefont {S.}~\bibnamefont
  {Bhattacharya}}, \bibinfo {author} {\bibfnamefont {P.}~\bibnamefont
  {Chandra}}, \bibinfo {author} {\bibfnamefont {V.}~\bibnamefont {Keppens}},
  \bibinfo {author} {\bibfnamefont {D.}~\bibnamefont {Mandrus}}, \bibinfo
  {author} {\bibfnamefont {M.~A.}\ \bibnamefont {McGuire}}, \bibinfo {author}
  {\bibfnamefont {B.~C.}\ \bibnamefont {Sales}}, \bibinfo {author}
  {\bibfnamefont {A.~S.}\ \bibnamefont {Sefat}}, \ and\ \bibinfo {author}
  {\bibfnamefont {J.}~\bibnamefont {Schmalian}},\ }\href {\doibase
  10.1103/PhysRevLett.105.157003} {\bibfield  {journal} {\bibinfo  {journal}
  {Phys. Rev. Lett.}\ }\textbf {\bibinfo {volume} {105}},\ \bibinfo {pages}
  {157003} (\bibinfo {year} {2010})}\BibitemShut {NoStop}%
\bibitem [{\citenamefont {Lee}\ \emph {et~al.}(2009)\citenamefont {Lee},
  \citenamefont {Yin},\ and\ \citenamefont {Ku}}]{CCLee2009}%
  \BibitemOpen
  \bibfield  {author} {\bibinfo {author} {\bibfnamefont {C.-C.}\ \bibnamefont
  {Lee}}, \bibinfo {author} {\bibfnamefont {W.-G.}\ \bibnamefont {Yin}}, \ and\
  \bibinfo {author} {\bibfnamefont {W.}~\bibnamefont {Ku}},\ }\href {\doibase
  10.1103/PhysRevLett.103.267001} {\bibfield  {journal} {\bibinfo  {journal}
  {Phys. Rev. Lett.}\ }\textbf {\bibinfo {volume} {103}},\ \bibinfo {pages}
  {267001} (\bibinfo {year} {2009})}\BibitemShut {NoStop}%
\bibitem [{\citenamefont {Yi}\ \emph {et~al.}(2011)\citenamefont {Yi},
  \citenamefont {Lu}, \citenamefont {Chu}, \citenamefont {Analytis},
  \citenamefont {Sorini}, \citenamefont {Kemper}, \citenamefont {Moritz},
  \citenamefont {Mo}, \citenamefont {Moore}, \citenamefont {Hashimoto},
  \citenamefont {Lee}, \citenamefont {Hussain}, \citenamefont {Devereaux},
  \citenamefont {Fisher},\ and\ \citenamefont {Shen}}]{Yi2011}%
  \BibitemOpen
  \bibfield  {author} {\bibinfo {author} {\bibfnamefont {M.}~\bibnamefont
  {Yi}}, \bibinfo {author} {\bibfnamefont {D.}~\bibnamefont {Lu}}, \bibinfo
  {author} {\bibfnamefont {J.-H.}\ \bibnamefont {Chu}}, \bibinfo {author}
  {\bibfnamefont {J.~G.}\ \bibnamefont {Analytis}}, \bibinfo {author}
  {\bibfnamefont {A.~P.}\ \bibnamefont {Sorini}}, \bibinfo {author}
  {\bibfnamefont {A.~F.}\ \bibnamefont {Kemper}}, \bibinfo {author}
  {\bibfnamefont {B.}~\bibnamefont {Moritz}}, \bibinfo {author} {\bibfnamefont
  {S.-K.}\ \bibnamefont {Mo}}, \bibinfo {author} {\bibfnamefont {R.~G.}\
  \bibnamefont {Moore}}, \bibinfo {author} {\bibfnamefont {M.}~\bibnamefont
  {Hashimoto}}, \bibinfo {author} {\bibfnamefont {W.-S.}\ \bibnamefont {Lee}},
  \bibinfo {author} {\bibfnamefont {Z.}~\bibnamefont {Hussain}}, \bibinfo
  {author} {\bibfnamefont {T.~P.}\ \bibnamefont {Devereaux}}, \bibinfo {author}
  {\bibfnamefont {I.~R.}\ \bibnamefont {Fisher}}, \ and\ \bibinfo {author}
  {\bibfnamefont {Z.-X.}\ \bibnamefont {Shen}},\ }\href {\doibase
  10.1073/pnas.1015572108} {\bibfield  {journal} {\bibinfo  {journal}
  {Proceedings of the National Academy of Sciences}\ }\textbf {\bibinfo
  {volume} {108}},\ \bibinfo {pages} {6878} (\bibinfo {year} {2011})},\ \Eprint
  {http://arxiv.org/abs/http://www.pnas.org/content/108/17/6878.full.pdf}
  {http://www.pnas.org/content/108/17/6878.full.pdf} \BibitemShut {NoStop}%
\bibitem [{\citenamefont {Liang}\ \emph {et~al.}(2013)\citenamefont {Liang},
  \citenamefont {Moreo},\ and\ \citenamefont {Dagotto}}]{Liang2013}%
  \BibitemOpen
  \bibfield  {author} {\bibinfo {author} {\bibfnamefont {S.}~\bibnamefont
  {Liang}}, \bibinfo {author} {\bibfnamefont {A.}~\bibnamefont {Moreo}}, \ and\
  \bibinfo {author} {\bibfnamefont {E.}~\bibnamefont {Dagotto}},\ }\href
  {\doibase 10.1103/PhysRevLett.111.047004} {\bibfield  {journal} {\bibinfo
  {journal} {Phys. Rev. Lett.}\ }\textbf {\bibinfo {volume} {111}},\ \bibinfo
  {pages} {047004} (\bibinfo {year} {2013})}\BibitemShut {NoStop}%
\bibitem [{\citenamefont {Fernandes}\ \emph {et~al.}(2014)\citenamefont
  {Fernandes}, \citenamefont {Chubukov},\ and\ \citenamefont
  {Schmalian}}]{Fernandes2014}%
  \BibitemOpen
  \bibfield  {author} {\bibinfo {author} {\bibfnamefont {R.~M.}\ \bibnamefont
  {Fernandes}}, \bibinfo {author} {\bibfnamefont {A.~V.}\ \bibnamefont
  {Chubukov}}, \ and\ \bibinfo {author} {\bibfnamefont {J.}~\bibnamefont
  {Schmalian}},\ }\href {http://dx.doi.org/10.1038/nphys2877} {\bibfield
  {journal} {\bibinfo  {journal} {Nat Phys}\ }\textbf {\bibinfo {volume}
  {10}},\ \bibinfo {pages} {97} (\bibinfo {year} {2014})},\ \bibinfo {note}
  {review}\BibitemShut {NoStop}%
\bibitem [{\citenamefont {Niedziela}\ \emph {et~al.}(2011)\citenamefont
  {Niedziela}, \citenamefont {Parshall}, \citenamefont {Lokshin}, \citenamefont
  {Sefat}, \citenamefont {Alatas},\ and\ \citenamefont
  {Egami}}]{Niedziela2011}%
  \BibitemOpen
  \bibfield  {author} {\bibinfo {author} {\bibfnamefont {J.~L.}\ \bibnamefont
  {Niedziela}}, \bibinfo {author} {\bibfnamefont {D.}~\bibnamefont {Parshall}},
  \bibinfo {author} {\bibfnamefont {K.~A.}\ \bibnamefont {Lokshin}}, \bibinfo
  {author} {\bibfnamefont {A.~S.}\ \bibnamefont {Sefat}}, \bibinfo {author}
  {\bibfnamefont {A.}~\bibnamefont {Alatas}}, \ and\ \bibinfo {author}
  {\bibfnamefont {T.}~\bibnamefont {Egami}},\ }\href {\doibase
  10.1103/PhysRevB.84.224305} {\bibfield  {journal} {\bibinfo  {journal} {Phys.
  Rev. B}\ }\textbf {\bibinfo {volume} {84}},\ \bibinfo {pages} {224305}
  (\bibinfo {year} {2011})}\BibitemShut {NoStop}%
\bibitem [{\citenamefont {Fernandes}\ \emph {et~al.}(2013)\citenamefont
  {Fernandes}, \citenamefont {B\"ohmer}, \citenamefont {Meingast},\ and\
  \citenamefont {Schmalian}}]{Fernandes2013}%
  \BibitemOpen
  \bibfield  {author} {\bibinfo {author} {\bibfnamefont {R.~M.}\ \bibnamefont
  {Fernandes}}, \bibinfo {author} {\bibfnamefont {A.~E.}\ \bibnamefont
  {B\"ohmer}}, \bibinfo {author} {\bibfnamefont {C.}~\bibnamefont {Meingast}},
  \ and\ \bibinfo {author} {\bibfnamefont {J.}~\bibnamefont {Schmalian}},\
  }\href {\doibase 10.1103/PhysRevLett.111.137001} {\bibfield  {journal}
  {\bibinfo  {journal} {Phys. Rev. Lett.}\ }\textbf {\bibinfo {volume} {111}},\
  \bibinfo {pages} {137001} (\bibinfo {year} {2013})}\BibitemShut {NoStop}%
\bibitem [{\citenamefont {Hardy}\ \emph {et~al.}(2010)\citenamefont {Hardy},
  \citenamefont {Wolf}, \citenamefont {Fisher}, \citenamefont {Eder},
  \citenamefont {Schweiss}, \citenamefont {Adelmann}, \citenamefont
  {v.~L\"ohneysen},\ and\ \citenamefont {Meingast}}]{Hardy2010}%
  \BibitemOpen
  \bibfield  {author} {\bibinfo {author} {\bibfnamefont {F.}~\bibnamefont
  {Hardy}}, \bibinfo {author} {\bibfnamefont {T.}~\bibnamefont {Wolf}},
  \bibinfo {author} {\bibfnamefont {R.~A.}\ \bibnamefont {Fisher}}, \bibinfo
  {author} {\bibfnamefont {R.}~\bibnamefont {Eder}}, \bibinfo {author}
  {\bibfnamefont {P.}~\bibnamefont {Schweiss}}, \bibinfo {author}
  {\bibfnamefont {P.}~\bibnamefont {Adelmann}}, \bibinfo {author}
  {\bibfnamefont {H.}~\bibnamefont {v.~L\"ohneysen}}, \ and\ \bibinfo {author}
  {\bibfnamefont {C.}~\bibnamefont {Meingast}},\ }\href {\doibase
  10.1103/PhysRevB.81.060501} {\bibfield  {journal} {\bibinfo  {journal} {Phys.
  Rev. B}\ }\textbf {\bibinfo {volume} {81}},\ \bibinfo {pages} {060501}
  (\bibinfo {year} {2010})}\BibitemShut {NoStop}%
\bibitem [{\citenamefont {Tegel}\ \emph {et~al.}(2008)\citenamefont {Tegel},
  \citenamefont {Johansson}, \citenamefont {Wei{\ss{}}}, \citenamefont
  {Schellenberg}, \citenamefont {Hermes}, \citenamefont {P{\"{o}}ttgen},\ and\
  \citenamefont {Johrendt}}]{Tegel08-1}%
  \BibitemOpen
  \bibfield  {author} {\bibinfo {author} {\bibfnamefont {M.}~\bibnamefont
  {Tegel}}, \bibinfo {author} {\bibfnamefont {S.}~\bibnamefont {Johansson}},
  \bibinfo {author} {\bibfnamefont {V.}~\bibnamefont {Wei{\ss{}}}}, \bibinfo
  {author} {\bibfnamefont {I.}~\bibnamefont {Schellenberg}}, \bibinfo {author}
  {\bibfnamefont {W.}~\bibnamefont {Hermes}}, \bibinfo {author} {\bibfnamefont
  {R.}~\bibnamefont {P{\"{o}}ttgen}}, \ and\ \bibinfo {author} {\bibfnamefont
  {D.}~\bibnamefont {Johrendt}},\ }\href {\doibase 10.1209/0295-5075/84/67007}
  {\bibfield  {journal} {\bibinfo  {journal} {EPL (Europhysics Letters)}\
  }\textbf {\bibinfo {volume} {84}},\ \bibinfo {pages} {67007} (\bibinfo {year}
  {2008})}\BibitemShut {NoStop}%
\bibitem [{\citenamefont {Rotter}\ \emph {et~al.}(2008)\citenamefont {Rotter},
  \citenamefont {Tegel}, \citenamefont {Johrendt}, \citenamefont
  {Schellenberg}, \citenamefont {Hermes},\ and\ \citenamefont
  {P\"ottgen}}]{Rotter08}%
  \BibitemOpen
  \bibfield  {author} {\bibinfo {author} {\bibfnamefont {M.}~\bibnamefont
  {Rotter}}, \bibinfo {author} {\bibfnamefont {M.}~\bibnamefont {Tegel}},
  \bibinfo {author} {\bibfnamefont {D.}~\bibnamefont {Johrendt}}, \bibinfo
  {author} {\bibfnamefont {I.}~\bibnamefont {Schellenberg}}, \bibinfo {author}
  {\bibfnamefont {W.}~\bibnamefont {Hermes}}, \ and\ \bibinfo {author}
  {\bibfnamefont {R.}~\bibnamefont {P\"ottgen}},\ }\href {\doibase
  10.1103/PhysRevB.78.020503} {\bibfield  {journal} {\bibinfo  {journal} {Phys.
  Rev. B}\ }\textbf {\bibinfo {volume} {78}},\ \bibinfo {pages} {020503}
  (\bibinfo {year} {2008})}\BibitemShut {NoStop}%
\bibitem [{\citenamefont {Ewings}\ \emph {et~al.}(2011)\citenamefont {Ewings},
  \citenamefont {Perring}, \citenamefont {Gillett}, \citenamefont {Das},
  \citenamefont {Sebastian}, \citenamefont {Taylor}, \citenamefont {Guidi},\
  and\ \citenamefont {Boothroyd}}]{Ewings2011}%
  \BibitemOpen
  \bibfield  {author} {\bibinfo {author} {\bibfnamefont {R.~A.}\ \bibnamefont
  {Ewings}}, \bibinfo {author} {\bibfnamefont {T.~G.}\ \bibnamefont {Perring}},
  \bibinfo {author} {\bibfnamefont {J.}~\bibnamefont {Gillett}}, \bibinfo
  {author} {\bibfnamefont {S.~D.}\ \bibnamefont {Das}}, \bibinfo {author}
  {\bibfnamefont {S.~E.}\ \bibnamefont {Sebastian}}, \bibinfo {author}
  {\bibfnamefont {A.~E.}\ \bibnamefont {Taylor}}, \bibinfo {author}
  {\bibfnamefont {T.}~\bibnamefont {Guidi}}, \ and\ \bibinfo {author}
  {\bibfnamefont {A.~T.}\ \bibnamefont {Boothroyd}},\ }\href {\doibase
  10.1103/PhysRevB.83.214519} {\bibfield  {journal} {\bibinfo  {journal} {Phys.
  Rev. B}\ }\textbf {\bibinfo {volume} {83}},\ \bibinfo {pages} {214519}
  (\bibinfo {year} {2011})}\BibitemShut {NoStop}%
\bibitem [{\citenamefont {Matan}\ \emph {et~al.}(2010)\citenamefont {Matan},
  \citenamefont {Ibuka}, \citenamefont {Morinaga}, \citenamefont {Chi},
  \citenamefont {Lynn}, \citenamefont {Christianson}, \citenamefont {Lumsden},\
  and\ \citenamefont {Sato}}]{Matan10}%
  \BibitemOpen
  \bibfield  {author} {\bibinfo {author} {\bibfnamefont {K.}~\bibnamefont
  {Matan}}, \bibinfo {author} {\bibfnamefont {S.}~\bibnamefont {Ibuka}},
  \bibinfo {author} {\bibfnamefont {R.}~\bibnamefont {Morinaga}}, \bibinfo
  {author} {\bibfnamefont {S.}~\bibnamefont {Chi}}, \bibinfo {author}
  {\bibfnamefont {J.~W.}\ \bibnamefont {Lynn}}, \bibinfo {author}
  {\bibfnamefont {A.~D.}\ \bibnamefont {Christianson}}, \bibinfo {author}
  {\bibfnamefont {M.~D.}\ \bibnamefont {Lumsden}}, \ and\ \bibinfo {author}
  {\bibfnamefont {T.~J.}\ \bibnamefont {Sato}},\ }\href {\doibase
  10.1103/PhysRevB.82.054515} {\bibfield  {journal} {\bibinfo  {journal} {Phys.
  Rev. B}\ }\textbf {\bibinfo {volume} {82}},\ \bibinfo {pages} {054515}
  (\bibinfo {year} {2010})}\BibitemShut {NoStop}%
\bibitem [{\citenamefont {Avci}\ \emph {et~al.}(2011)\citenamefont {Avci},
  \citenamefont {Chmaissem}, \citenamefont {Goremychkin}, \citenamefont
  {Rosenkranz}, \citenamefont {Castellan}, \citenamefont {Chung}, \citenamefont
  {Todorov}, \citenamefont {Schlueter}, \citenamefont {Claus}, \citenamefont
  {Kanatzidis}, \citenamefont {Daoud-Aladine}, \citenamefont {Khalyavin},\ and\
  \citenamefont {Osborn}}]{Avci2011}%
  \BibitemOpen
  \bibfield  {author} {\bibinfo {author} {\bibfnamefont {S.}~\bibnamefont
  {Avci}}, \bibinfo {author} {\bibfnamefont {O.}~\bibnamefont {Chmaissem}},
  \bibinfo {author} {\bibfnamefont {E.~A.}\ \bibnamefont {Goremychkin}},
  \bibinfo {author} {\bibfnamefont {S.}~\bibnamefont {Rosenkranz}}, \bibinfo
  {author} {\bibfnamefont {J.-P.}\ \bibnamefont {Castellan}}, \bibinfo {author}
  {\bibfnamefont {D.~Y.}\ \bibnamefont {Chung}}, \bibinfo {author}
  {\bibfnamefont {I.~S.}\ \bibnamefont {Todorov}}, \bibinfo {author}
  {\bibfnamefont {J.~A.}\ \bibnamefont {Schlueter}}, \bibinfo {author}
  {\bibfnamefont {H.}~\bibnamefont {Claus}}, \bibinfo {author} {\bibfnamefont
  {M.~G.}\ \bibnamefont {Kanatzidis}}, \bibinfo {author} {\bibfnamefont
  {A.}~\bibnamefont {Daoud-Aladine}}, \bibinfo {author} {\bibfnamefont
  {D.}~\bibnamefont {Khalyavin}}, \ and\ \bibinfo {author} {\bibfnamefont
  {R.}~\bibnamefont {Osborn}},\ }\href {\doibase 10.1103/PhysRevB.83.172503}
  {\bibfield  {journal} {\bibinfo  {journal} {Phys. Rev. B}\ }\textbf {\bibinfo
  {volume} {83}},\ \bibinfo {pages} {172503} (\bibinfo {year}
  {2011})}\BibitemShut {NoStop}%
\bibitem [{\citenamefont {Cano}\ \emph {et~al.}(2010)\citenamefont {Cano},
  \citenamefont {Civelli}, \citenamefont {Eremin},\ and\ \citenamefont
  {Paul}}]{Cano2010}%
  \BibitemOpen
  \bibfield  {author} {\bibinfo {author} {\bibfnamefont {A.}~\bibnamefont
  {Cano}}, \bibinfo {author} {\bibfnamefont {M.}~\bibnamefont {Civelli}},
  \bibinfo {author} {\bibfnamefont {I.}~\bibnamefont {Eremin}}, \ and\ \bibinfo
  {author} {\bibfnamefont {I.}~\bibnamefont {Paul}},\ }\href {\doibase
  10.1103/PhysRevB.82.020408} {\bibfield  {journal} {\bibinfo  {journal} {Phys.
  Rev. B}\ }\textbf {\bibinfo {volume} {82}},\ \bibinfo {pages} {020408}
  (\bibinfo {year} {2010})}\BibitemShut {NoStop}%
\bibitem [{\citenamefont {Chu}\ \emph {et~al.}(2010)\citenamefont {Chu},
  \citenamefont {Analytis}, \citenamefont {De~Greve}, \citenamefont {McMahon},
  \citenamefont {Islam}, \citenamefont {Yamamoto},\ and\ \citenamefont
  {Fisher}}]{ChuJH2010}%
  \BibitemOpen
  \bibfield  {author} {\bibinfo {author} {\bibfnamefont {J.-H.}\ \bibnamefont
  {Chu}}, \bibinfo {author} {\bibfnamefont {J.~G.}\ \bibnamefont {Analytis}},
  \bibinfo {author} {\bibfnamefont {K.}~\bibnamefont {De~Greve}}, \bibinfo
  {author} {\bibfnamefont {P.~L.}\ \bibnamefont {McMahon}}, \bibinfo {author}
  {\bibfnamefont {Z.}~\bibnamefont {Islam}}, \bibinfo {author} {\bibfnamefont
  {Y.}~\bibnamefont {Yamamoto}}, \ and\ \bibinfo {author} {\bibfnamefont
  {I.~R.}\ \bibnamefont {Fisher}},\ }\href {\doibase 10.1126/science.1190482}
  {\bibfield  {journal} {\bibinfo  {journal} {Science}\ }\textbf {\bibinfo
  {volume} {329}},\ \bibinfo {pages} {824} (\bibinfo {year} {2010})},\ \Eprint
  {http://arxiv.org/abs/http://www.sciencemag.org/content/329/5993/824.full.pdf}
  {http://www.sciencemag.org/content/329/5993/824.full.pdf} \BibitemShut
  {NoStop}%
\bibitem [{\citenamefont {Parshall}\ \emph {et~al.}(2014)\citenamefont
  {Parshall}, \citenamefont {Heid}, \citenamefont {Niedziela}, \citenamefont
  {Wolf}, \citenamefont {Stone}, \citenamefont {Abernathy},\ and\ \citenamefont
  {Reznik}}]{Parshall2014}%
  \BibitemOpen
  \bibfield  {author} {\bibinfo {author} {\bibfnamefont {D.}~\bibnamefont
  {Parshall}}, \bibinfo {author} {\bibfnamefont {R.}~\bibnamefont {Heid}},
  \bibinfo {author} {\bibfnamefont {J.~L.}\ \bibnamefont {Niedziela}}, \bibinfo
  {author} {\bibfnamefont {T.}~\bibnamefont {Wolf}}, \bibinfo {author}
  {\bibfnamefont {M.~B.}\ \bibnamefont {Stone}}, \bibinfo {author}
  {\bibfnamefont {D.~L.}\ \bibnamefont {Abernathy}}, \ and\ \bibinfo {author}
  {\bibfnamefont {D.}~\bibnamefont {Reznik}},\ }\href {\doibase
  10.1103/PhysRevB.89.064310} {\bibfield  {journal} {\bibinfo  {journal} {Phys.
  Rev. B}\ }\textbf {\bibinfo {volume} {89}},\ \bibinfo {pages} {064310}
  (\bibinfo {year} {2014})}\BibitemShut {NoStop}%
\bibitem [{\citenamefont {Boeri}\ \emph {et~al.}(2010)\citenamefont {Boeri},
  \citenamefont {Calandra}, \citenamefont {Mazin}, \citenamefont {Dolgov},\
  and\ \citenamefont {Mauri}}]{Boeri2010}%
  \BibitemOpen
  \bibfield  {author} {\bibinfo {author} {\bibfnamefont {L.}~\bibnamefont
  {Boeri}}, \bibinfo {author} {\bibfnamefont {M.}~\bibnamefont {Calandra}},
  \bibinfo {author} {\bibfnamefont {I.~I.}\ \bibnamefont {Mazin}}, \bibinfo
  {author} {\bibfnamefont {O.~V.}\ \bibnamefont {Dolgov}}, \ and\ \bibinfo
  {author} {\bibfnamefont {F.}~\bibnamefont {Mauri}},\ }\href {\doibase
  10.1103/PhysRevB.82.020506} {\bibfield  {journal} {\bibinfo  {journal} {Phys.
  Rev. B}\ }\textbf {\bibinfo {volume} {82}},\ \bibinfo {pages} {020506}
  (\bibinfo {year} {2010})}\BibitemShut {NoStop}%
\bibitem [{\citenamefont {Mazin}\ \emph {et~al.}(2008)\citenamefont {Mazin},
  \citenamefont {Johannes}, \citenamefont {Boeri}, \citenamefont {Koepernik},\
  and\ \citenamefont {Singh}}]{Mazin08-1}%
  \BibitemOpen
  \bibfield  {author} {\bibinfo {author} {\bibfnamefont {I.~I.}\ \bibnamefont
  {Mazin}}, \bibinfo {author} {\bibfnamefont {M.~D.}\ \bibnamefont {Johannes}},
  \bibinfo {author} {\bibfnamefont {L.}~\bibnamefont {Boeri}}, \bibinfo
  {author} {\bibfnamefont {K.}~\bibnamefont {Koepernik}}, \ and\ \bibinfo
  {author} {\bibfnamefont {D.~J.}\ \bibnamefont {Singh}},\ }\href {\doibase
  10.1103/PhysRevB.78.085104} {\bibfield  {journal} {\bibinfo  {journal} {Phys.
  Rev. B}\ }\textbf {\bibinfo {volume} {78}},\ \bibinfo {pages} {085104}
  (\bibinfo {year} {2008})}\BibitemShut {NoStop}%
\end{thebibliography}
\end{document}